\documentclass[11pt,twoside]{article}

\usepackage{asp2004}
\usepackage{epsf}
\usepackage{psfig}
\usepackage{lscape}
\usepackage{graphicx}
\usepackage{amsmath}
\usepackage{amssymb}

\newcommand{\ez}{{\rm {\bf e_{\rm z}}}}

\def\depp{^{*}}
\def\deppp{^{2*}}
\newcommand{\ve}[1]{{\rm{\bf {#1}}}}

\begin{document}
\title{The uncombed penumbra}   
\author{J.M.~Borrero \& M. Rempel}   
\affil{High Altitude Observatory, National Corporation for
Atmospheric Research, 3450 Mitchell Lane 80301, Boulder 80301 Colorado USA.}    
\author{S.K.~Solanki}
\affil{Max Planck Institut f\"ur Sonnensystemforschung, 37191 Katlenburg-Lindau,
Germany}


\begin{abstract} 
The uncombed penumbral model explains the
structure of the sunspot penumbra in terms of thick magnetic fibrils
embedded in a magnetic surrounding atmosphere. This model
has been successfully applied to explain the polarization
signals emerging from the sunspot penumbra. Thick penumbral 
fibrils face some physical problems, however. In this contribution
we will offer possible solutions to these shortcomings.
\end{abstract}

\section{Introduction}%

The structure of the penumbra has been a subject of
intensive research in the last years. Most of
our current knowledge is based on the interpretation
of the polarized line profiles that carry useful
information about the magnetic field topology.
A consistent picture of the sunspot penumbra, 
able to explain the various observations available
at different wavelength ranges, different spatial resolutions,
etc, has not yet emerged (Solanki 2003).

A widely used model is the so-called {\it uncombed penumbral model} by
Solanki \& Montavon (1993). It is based on the idea of a penumbra 
consisting of highly inclined magnetic flux tubes\footnote{In this 
paper we will use indistinctly the term magnetic fibril or magnetic
flux tube. Note that as a matter of fact the fibrils
correspond to an anti-flux tube since its magnetic field
strength is smaller that in the magnetic surrounding atmosphere 
(see Fig.~6 in Borrero et al. 2005)} embedded in a more vertical
magnetic background (for practical implementations a 
simplified version is used: see Borrero et al. 2003, 
2005). This model has been specially successful 
in explaining a number of key observations:

\begin{itemize}
\item it reproduces the properties of the Net Circular Polarization 
(magnitude, sign, distribution and center-to-limb variation)
observed in the sunspot penumbra in the visible Fe I lines at 
6300 \AA~ (Solanki \& Montavon 1993; Mart{\'\i}nez Pillet 2000) and 
Fe I lines at 1.56 $\mu$m (Schlichenmaier \& Collados 2002; Schlichenmaier 
et al. 2002; M\"uller et al. 2002).
\item it offers an explanation for the opposite vertical gradients (in the 
line-of-sight velocity, magnetic field inclination and magnetic field 
strength) obtained from the inversion of spectropolarimetric data 
of the spectral lines when different spectral lines are used (Westendorp
Plaza et al. 2001a, 2001b; Mathew et al. 2003; Borrero et al. 2004).
\item it consistently reproduces the polarization signals
emerging from the sunspot penumbra in a variety of spectral lines
and sunspots at different heliocentric angles (Borrero et al. 2005; 
Borrero et al. 2006).
\item it retrieves flux tubes whose vertical extension
is about 100-300 km (see left panel in Fig.~1). 
This is comparable to the horizontal 
extension seen in high resolution continuum images (Scharmer et al.
2002; Rouppe van der Voort et al. 2004; S\"utterlin et al. 2004). 
\end{itemize}

In addition to this, numerical simulations of thin penumbral flux 
tubes are able to explain the proper movements of the penumbral 
grains and moving magnetic features (Schlichenmaier 2002), 
as well as various features of the Evershed flow (Thomas \& Montesinos 1993; 
Montesinos \& Thomas 1997; Schlichenmaier et al. 1998a, 1998b). 

The magnetic topology inferred from the application
of the uncombed model to spectropolarimetric observations
is very similar to that used in numerical simulations
of penumbral flux tubes. The main difference lies in the vertical
extension of the penumbral fibrils. While numerical simulations
consider those flux tubes to be thin (much smaller than the
typical pressure scale height), spectropolarimetric observations
indicate that this might not be the case. Note that from
the inversion of Stokes profiles it is not possible to distinguish 
between a thick flux tube or a bundle of thin flux tubes next 
to each other (see discussion in Borrero et al. 2006).

In the {\it thin} case, numerical simulations would have
problems offering an explanation for the heating and brightness 
of the penumbra (Schlichenmaier \& Solanki 2003; Spruit \& Scharmer 2006). 
Even if the flux tubes carry hot plasma at say, 12000 K, its cools down 
so fast (Schlichenmaier et al. 1999) that the only possibility left is
either that the flux tube is thick (in order to increase the cooling time) 
or there are many thin flux tubes per resolution element that carry hot 
plasma upflows into the penumbra, cool down and sink again into deeper layers
within the same resolution element. Rouppe van der Voort et al. (2004)
and Langhans et al. (2005) observe long lived ($\sim$ 1 hr) 
penumbral filaments that are highly coherent over portions of the penumbra 
of several thousand kilometers. This seems to rule out the last possibility. 
We therefore want to turn our attention on the {\it thick} case: flux tubes 
of 100-300 km diameter.

\begin{figure}[!ht]
\begin{center}
\begin{tabular}{cc}
\includegraphics[width=6.5cm]{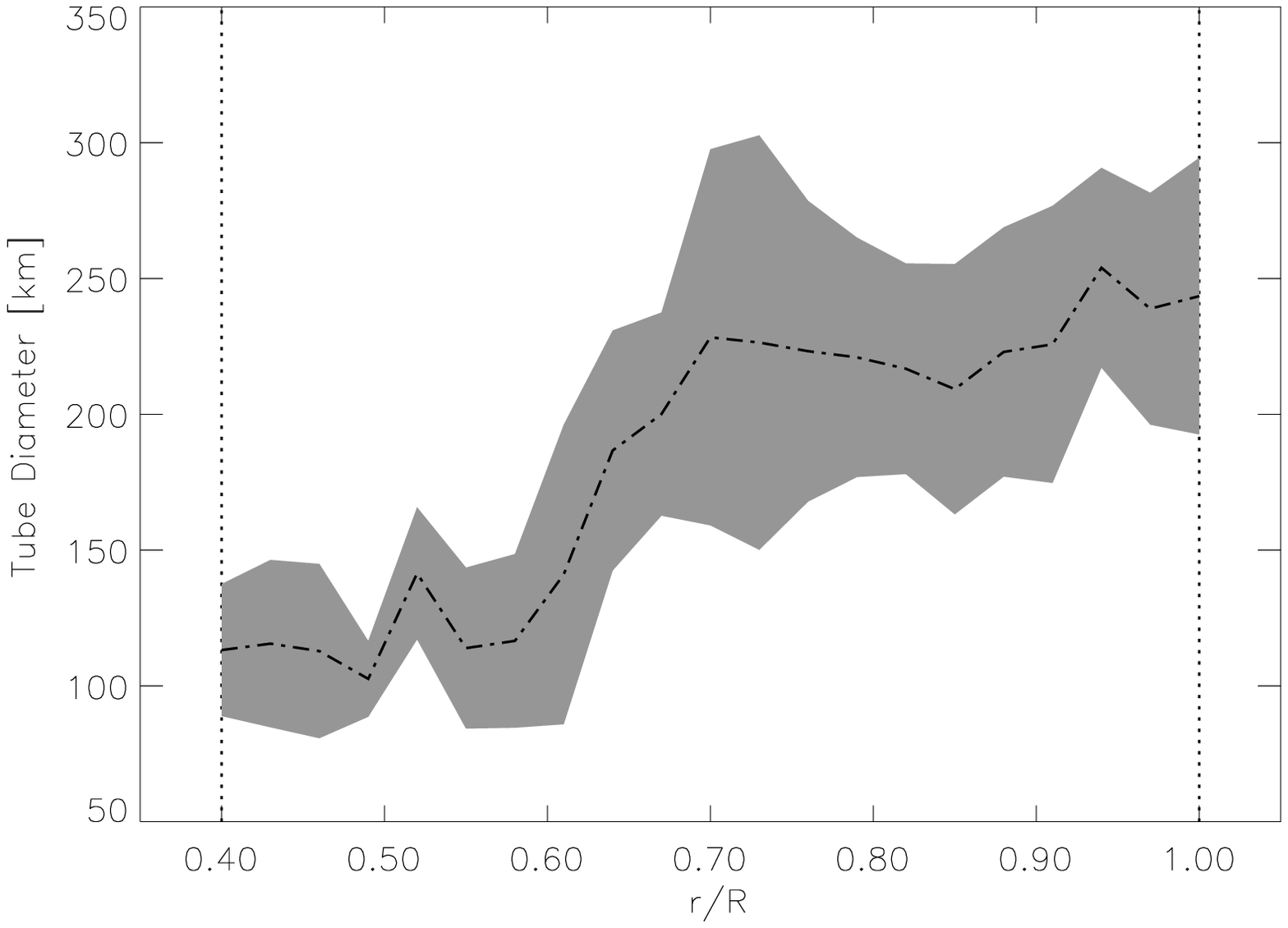} &
\includegraphics[width=6.5cm]{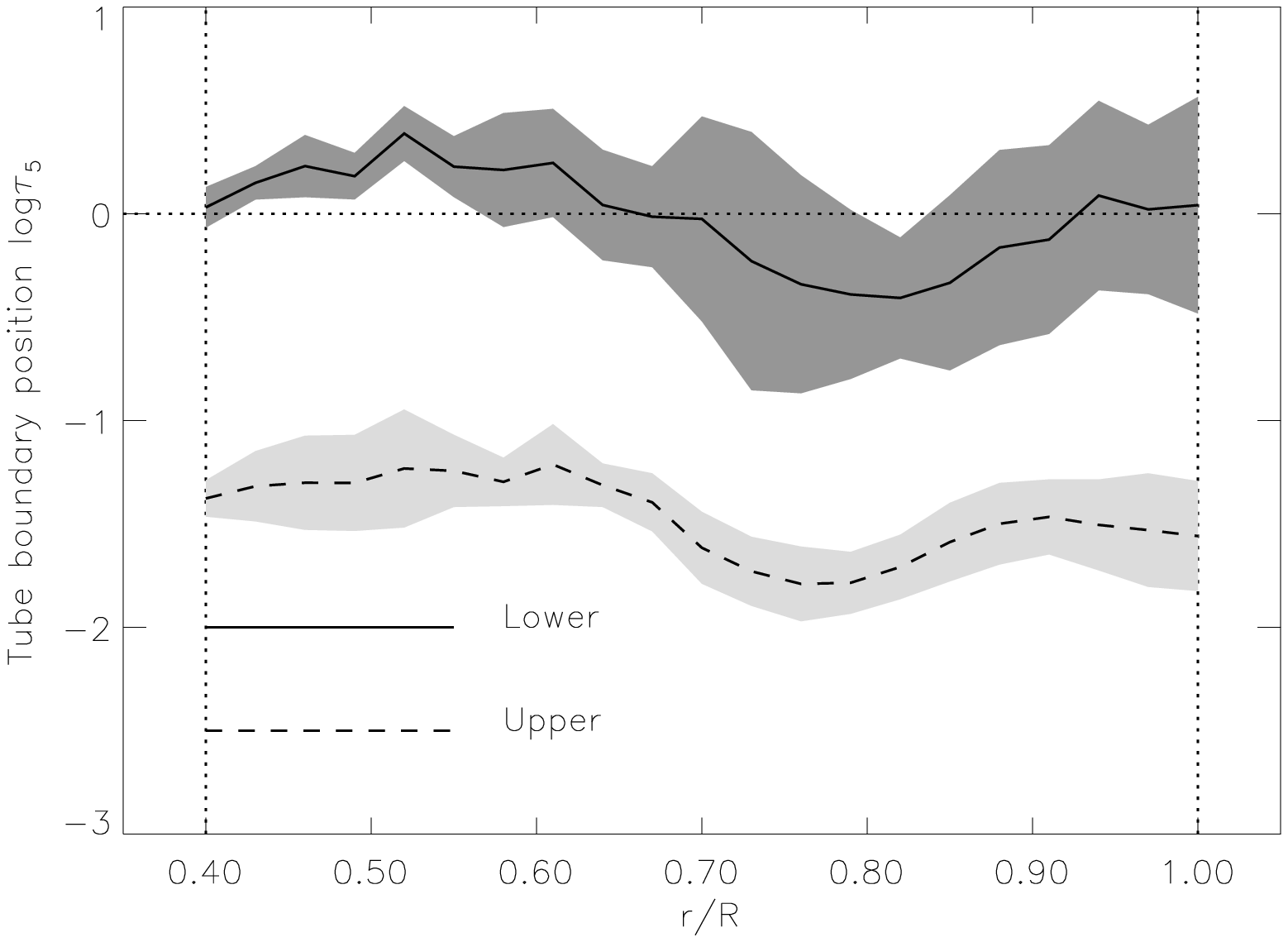}
\end{tabular}
\end{center}
\caption{{\it Left panel}: flux tube diameter as a function 
of the sunspot normalized radial distance in the penumbra.
{\it Right panel}: radial variation of the lower (solid line) 
and upper (dashed) flux tube's boundaries in the optical
depth scale. Both figures from Borrero et al. 2006.}
\end{figure}

\section{Problems in thick embedded flux tubes}%

\begin{figure}[!ht]
\begin{center}
\includegraphics[width=8cm]{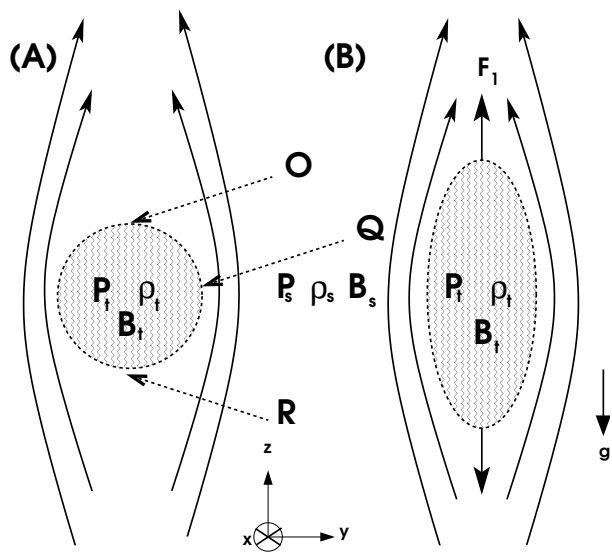}
\end{center}
\caption{{\it Panel A (left)}: a homogeneous flux tube
with density $\rho_t$ and pressure $P_t$ and horizontal (with
respect to the vertical in the solar surface) magnetic field
$\ve{B_t}=(B_{tx},0,0)$, is place in a homogeneous environment
with $\rho_s$ and $P_s$ where the magnetic field is vertical:
$\ve{B_s}=(0,B_{sy},B_{sz})$. The external field lines bend around the flux 
tube, creating tension forces that stretch the flux tube vertically
({\it right panel B}).}
\end{figure}

When a horizontal (initially homogeneous) magnetic flux tube $\ve{B_t}$ is 
embedded in an external vertical magnetic field $\ve{B_s}$, this external 
or surrounding field has to bend aside in order to accommodate the flux tube. 
A situation like this is depicted in Fig.~2 (panel A). The component of the
magnetic field vector along the direction perpendicular to the flux tube's
surface vanishes for both the flux tube and surrounding field. This leads
to total pressure balance between them:

\begin{equation}
P_t\depp + \frac{B_t\deppp}{8\pi} = P_s\depp + \frac{B_s\deppp}{8\pi}
\end{equation}

\noindent where the symbol $\depp$ indicates the boundary or interface
between the flux tube and the magnetic surrounding atmosphere. On the
laterals of the flux tubes (point {\bf Q} in Fig.~2) we have:

\begin{equation}
P_t-P_s = \frac{B_{sz}^2-B_{tx}^2}{8\pi}
\end{equation}

while on top and at the bottom of the flux tube (points {\bf O} and points {\bf R} 
in Fig.~2) since the external magnetic field vanishes completely we have:

\begin{equation}
P_t-P_s = - \frac{B_{tx}^2}{8\pi}
\end{equation}

Any $P_t-P_s$ that balances Eq.~2 unbalances Eq.~3 in
the amount of $B_{sz}^2/8\pi$. This imbalance induces
net forces at the top and the bottom the flux tube that tend 
to stretch it vertically:

\begin{eqnarray}
\ve{F}_1^{O} & = & \frac{B_{sz}^2}{8\pi R}\ez \\
\ve{F}_1^{R} & = & - \frac{B_{sz}^2}{8\pi R}\ez
\end{eqnarray}

These forces causes the flux tube to expand vertically in a time scale of

\begin{equation}
t_{\rm expan} = \frac{l}{v} \sim \frac{R}{v_a} \sim \frac{R}{B}\sqrt{4\pi\rho}
\sim 20 \;\; {\rm s}
\end{equation}

\noindent where we assumed a flux tube radius of $R=100$ km,
a typical density of $\rho = 3\times 10^{-7}$ g cm$^{-3}$ and
a field strength of $B=1000$ Gauss.

If nothing stops this process, the flux tube upper boundary will indefinitely
move upwards and the bottom one will move downwards (see panel B in Fig.~2). 
This will end up breaking our idea of flux tube and of uncombed penumbra, since the
gradients at flux tube's boundaries (needed in the uncombed model to create NCP) 
will disappear from the regions in which the spectral lines are formed.
In the following we will consider if this stretching process might be kept 
within limits.

\section{Flux tubes in convectively stable layers}%

Let us assume that, as a consequence of the vertical stretching
anticipated in the previous section, the upper part of our initially 
homogeneous flux tube rises from a height $z_0$ to 
a height $z$.  If the atmosphere is subadiabatic (superadibatic index
$\delta <0$), a restoring force will try to bring back the flux tube 
to its original position $z_0$:

\begin{eqnarray}
\ve{F}_2^{O} & = & \frac{\rho g \delta \Delta z_o}{H_p} \ez
\end{eqnarray}

\noindent where $H_p=P/\rho g$ is the pressure scale height.
Note that $\ve{F}_{2}^{O}$ opposes to $\ve{F}_{1}^{O}$ (Eq.~4)
since $\Delta z > 0$ and $\delta < 0$. The main question 
that rises now is: how much do the upper portions
of the flux tube rise before the anti-buoyant restoring force
compensates the magnetic forces ? This can be estimated by 
comparing Eq.~4 and Eq.~7:

\begin{equation}
\frac{\Delta z}{H_p} = - \frac{B_{sz}^2}{8\pi R \rho g}\frac{1}{\delta}
\sim -\frac{0.75}{\delta}
\end{equation}

Using a $\delta = -0.4$ (ideal gas at a constant temperature; see 
Moreno-Insertis \& Spruit 1989) and $B_{sz}=1250$ Gauss
we deduce that the vertical stretching occurring to the upper region of the flux tube
is of the order of one or two pressure scale heights $\Delta z \sim H_p 
= P/\rho g \sim 100-200$ km. At this height its density will be $1-\delta$ times 
the density of the surrounding magnetic atmosphere.\\

This value for the vertical stretching is at the 
limits of what is needed to keep the flux tube boundary within the spectral 
line forming region. Futhermore, these estimates have been done for the case
of the top and bottom of the flux tube. As we move towards the 
sides (point {\bf Q} is Fig.~2) the vertical stretching is would be smaller.

Similarly we can argue that the stretching of the lower portions
of the flux tube will not grow exponentially if these are
less dense that the surrounding atmosphere (since $\Delta z < 0$).
When both conditions (upper and lower boundaries of the flux tube) 
are brought together, we end up with a non homogeneous flux tube.
However, the lower boundary of the flux tube appears to be located near the
continuum forming layers (Fig.~1; right panel), where the assumption 
of $\delta < 0$ is not completely justified.\\

Note that $B_{sz}$ in Eq.~8 decreases strongly towards the outer penumbra,
where the external field becomes weak and inclined $B_{sz} \sim 650$ and therefore
$\Delta z/ H_p \sim -0.2/\delta$. Near the umbra however, 
$B_{sz}$ reaches values as high as $1700$ Gauss and the vertical
stretching is so large that the flux tube is likely to break disappear.
High spatial resolution images of the sunspot penumbra reveal that often, 
when the penumbral filaments pour into the umbra, they break into two individual 
filaments that continue moving inwards and form umbral dots.
The details presented here offer a scenario that might account
for this process.

\section{Conclusions}%

We have shown, using simple estimates that in a convectively stable
atmosphere the vertical stretching of horizontal flux tubes,
embedded in a penumbral field, is limited by buoyancy. It would be of
considerable interest to use numerical simulations to confirm that
the vertical stretching is compatible with the uncombed penumbral model.

\acknowledgements Thanks to Rolf Schlichenmaier and Tom Bogdan for 
useful discussions.



\end{document}